\documentclass[12pt]{article}
\usepackage[dvips]{graphicx}

\newcommand{\cd}{\makebox[0.08cm]{$\cdot$}}
\newcommand{\VEV}[1]{\left\langle{#1}\right\rangle}

\renewcommand{\bar}{\overline}

\begin{document}

\begin{flushright}
SLAC-PUB-10068 \\
JLAB-THY-03-35\\
UMN-D-03-4 \\
November 2003
\end{flushright}

\begin{center}
{{\bf\LARGE The covariant structure of light-front \\[.2ex]
wave functions and the behavior  \\[1ex]
of hadronic form factors\footnote{Work supported in part by
Department of Energy contracts DE-AC03-76SF00515,
DE-AC05-84ER40150, DE-FG02-98ER41087, by the LG Yonam Foundation
and by the Exchange Visitor Program P-1-00162.}}}

\bigskip
\bigskip
{\it Stanley J. Brodsky \\
Stanford Linear Accelerator Center \\
Stanford University, Stanford, California 94309 \\
and the
Thomas Jefferson National Accelerator Laboratory\\
Newport News, Virginia 23606, USA\\
E-mail:  sjbth@slac.stanford.edu}
\medskip

{\it John R. Hiller \\
Department of Physics \\
University of Minnesota-Duluth \\
 Duluth, Minnesota 55812\\
 E-mail: jhiller@d.umn.edu}
 \medskip

{\it Dae Sung Hwang \\
Department of Physics \\
Sejong University \\
Seoul 143--747, Korea\\
E-mail: dshwang@sejong.ac.kr}
\medskip

{\it Vladimir A. Karmanov \\ Lebedev Physical Institute \\
Leninsky prospect 53 \\ 119991  Moscow, Russia \\ E-mail:
karmanov@sci.lebedev.ru}
\end{center}
\vfill \eject
\begin{center}
{\bf\large Abstract}
\end{center}

We study the analytic structure of light-front wave functions (LFWFs) and its
consequences for hadron form factors using an explicitly Lorentz-invariant
formulation of the front form. The normal to the light front is specified by a
general null vector $\omega^\mu.$ The LFWFs with definite total angular momentum
are eigenstates of a {\it kinematic} angular momentum operator and satisfy all
Lorentz symmetries. They are analytic functions of the invariant mass squared of
the constituents $M^2_0= (\sum k^\mu)^2$ and the light-cone momentum fractions
$x_i= {k_i\cd \omega / p \cd \omega}$  multiplied by invariants constructed from
the spin matrices,  polarization vectors, and $\omega^\mu.$  These properties are
illustrated using known nonperturbative eigensolutions of the Wick--Cutkosky
model. We analyze the LFWFs introduced by Chung and Coester to describe static
and low momentum properties of the nucleons. They correspond to the spin-locking
of a quark with the spin of its parent nucleon, together with a positive-energy
projection constraint. These extra constraints lead to anomalous dependence of
form factors on $Q$ rather than $Q^2.$ In contrast, the dependence of LFWFs on
$M^2_0$ implies that hadron form factors are analytic functions of $Q^2$ in
agreement with dispersion theory and perturbative QCD. We show that a model
incorporating the leading-twist perturbative QCD prediction is consistent with
recent data for the ratio of proton Pauli and Dirac form factors.


\bigskip

\section{Introduction}

Light-front wave functions (LFWFs) are the interpolating functions
connecting hadrons to their fundamental quark and gluon degrees of
freedom in QCD.  Many hadronic observables can be computed
directly from these amplitudes.  For example, matrix elements of
local operators such as spacelike proton form factors, transition
form factors such as $B \to \ell \bar \nu \pi$, and generalized
parton distributions can be computed  from the overlap integrals
of the LFWFs.  The determination of the hadron LFWFs from
phenomenological constraints and from QCD itself is thus a central
goal of hadron and nuclear physics.   In principle, one can solve
for the hadronic LFWFs directly from fundamental theory using
nonperturbative methods, such as discretized light-front
quantization, the transverse lattice, lattice gauge theory
moments, or Bethe--Salpeter/Dyson-Schwinger techniques.   Reviews
of nonperturbative light-front methods may be found in
Refs.~\cite{cdkm,Brodsky:1997de,Dalley:ug}.

One of the central issues in the analysis of fundamental hadron
structure is the presence of nonzero orbital angular momentum in
the bound-state wave functions.  The evidence for a ``spin crisis"
in the Ellis-Jaffe sum rule signals a significant orbital
contribution in the proton wave
function~\cite{Jaffe:1989jz,Ji:2002qa}.  The Pauli form factor of
nucleons is computed from the overlap of LFWFs differing by one
unit of orbital angular momentum $\Delta L_z= \pm 1$.  Thus the
fact that the anomalous magnetic moment of the proton is not zero
is an immediate signal for the presence of nonzero orbital angular
momentum in the proton's LFWFs~\cite{BD80}.  It should be noted
that orbital angular momentum is treated explicitly in light-front
quantization; it includes the orbital contributions induced by
relativistic effects, such as the spin-orbit effects normally
associated with the conventional Dirac spinors.

In this paper we shall show how orbital angular momentum is represented by
light-front Fock state wave functions.  A key tool will be the explicitly
Lorentz-invariant formulation of the front form (see~\cite{cdkm} for a review and
references to original papers). The wave functions are defined at the light-front
plane $\omega\cd x=\sigma$, for which the orientation is determined by the null
four-vector $\omega$. Although LFWFs depend on the choice of the light-front
quantization direction $\omega$, all observables such as matrix elements of local
current operators, form factors, and cross sections are light-front invariants --
they must be $\omega$-independent. When the $\omega$-independence is violated in
approximate calculations, one can still find the $\omega$-independent form
factors by separating them from the current operator matrix elements and omitting
the non-physical, $\omega$-dependent contribution~\cite{km96}. We shall show that
the analytic form of LFWFs with nonzero orbital angular momentum is then
constrained to a specific set of simple prefactors multiplying the scalar zero
orbital angular momentum solutions. Knowing the general form of the LFWFs can be
important for determining the hadron eigenstates from QCD using variational or
other methods.

Our results for the analytic form of hadronic LFWFs, including orbital angular
momentum, are consistent at large transverse momentum with the perturbative QCD
counting rules of Ji, Ma, and Yuan~\cite{jmy} and the wavefunction
constraints~\cite{Brodsky:2003px} which follow the conformal properties of the
AdS/CFT correspondence between gauge theory and string
theory~\cite{Polchinski:2001tt,Maldacena:1997re}.

We begin by noting that eigensolutions of the Bethe-Salpeter
equation have specific angular momentum as specified by the
Pauli-Lubanski vector.  The corresponding LFWF for an $n$-particle
Fock state evaluated at equal light-front time $\sigma = \omega\cd
x$ can be obtained by integrating the Bethe-Salpeter solutions
over the corresponding relative light-front energies.  The
resulting LFWFs $\psi^I_n(x_i, k_{\perp i})$ are functions of the
light-cone momentum fractions $x_i$  and the invariant mass
squared of the constituents $M_0^2= (\sum^n_{i=1} k_i^\mu)^2
=\sum_{i =1}^n \big [{k^2_\perp + m^2\over x}\big]_i$ and the
light-cone momentum fractions $x_i= {k\cd \omega / p \cd \omega}$,
each multiplying spin-vector and polarization tensor invariants
which can involve $\omega^\mu.$ The resulting LFWFs for bound
states are eigenstates of a kinematic angular momentum operator.
Thus LFWFs satisfy all Lorentz symmetries of the front
form~\cite{Dirac:cp}, including boost invariance, and they are
proper eigenstates of angular momentum.

There is now heightened interest in the analytic form of the nucleon
form factors.  Recent measurements of the proton form factors at Jefferson
Laboratory~\cite{Jones:1999rz,Gayou:2001qd}, using the polarization
transfer method, show a surprising result -- the ratio \\
$G_E(Q^2)/G_M(Q^2)$ falls faster in momentum transfer $Q^2 = -q^2=
-t$ than that found using the traditional Rosenbluth separation
method.  A possible source for this disparity are the QED
radiative corrections, since these are more likely to affect the
Rosenbluth method~\cite{Blunden:2003sp,Guichon:2003qm,Afan}.  For
example, the interference of one-photon and two-photon exchange
amplitudes and the interference between proton and electron
bremsstrahlung are present in the measured electron-proton cross
section and can complicate the analysis of the energy and angular
dependence required for the Rosenbluth separation.

If one translates the new polarization transfer results for $G_E$
and $G_M$ to the Pauli and Dirac form factors, the data appear to
suggest the asymptotic behavior ${Q F_1(Q^2)/ F_2(Q^2)} \sim {\rm
const}$.  In a recent paper, Miller and Frank~\cite{mf} have shown
that the three-quark model for the proton LFWF constructed by
Chung and Coester (CC)~\cite{cc} and extended by
Schlumpf~\cite{Schlumpf} leads to ${Q F_1(Q^2)/ F_2(Q^2)} \sim
{\rm const}$  in the range of the JLab experiment, thus providing
an apparent explanation of the JLab data.

In dispersion theory form factors are analytic functions of $q^2$,
with a cut structure reflecting physical thresholds at timelike $q^2.$
This  is also apparent from the analytic structure of Feynman amplitudes
in perturbation theory.  A functional dependence in $Q=\sqrt{-q^2}$
does arise when there is a physical threshold at $q^2=0,$ as in the case of
gravitational~\cite{Brodsky:2000ii,Bjerrum-Bohr:2002kt} (or axial current)
form factors due to the two-photon intermediate state; however, this would
not be expected to occur for the vector current in QCD.

Chung and Coester~\cite{cc} introduced their ansatz for the form
of baryon LFWFs in order to describe the static and low momentum
transfer properties of the nucleons. The LFWFs in the CC model
have the effect of spin-locking a quark with the spin of its
parent nucleon, together with a positive-energy projection
constraint.  As we shall show, if one extends these forms to large
transverse momentum, the extra constraints lead to an anomalous
linear dependence of LFWFs in the invariant mass of the
constituents and an anomalous dependence  of form factors on $Q$
rather than $Q^2.$ As we discuss in the conclusions, the lack of
analyticity in $Q^2$ is related to the breakdown of the crossing
properties incorporated in field theory.  The CC constraint may
provide a reasonable  model for computing static properties of
hadrons, but it is not applicable to large momentum transfer
observables.

We shall show that form factors computed from the overlap of LFWFs
are analytic functions of $q^2$ due to their analytic dependence
on the off-shell light-front energy and the general form of
prefactors associated with nonzero orbital angular momentum.  In
particular, the  general form of the LFWFs for baryons in QCD
leads to a ratio of form factors $F_2(Q^2)/F_1(Q^2)$ which behaves
asymptotically as an inverse power of $Q^2$ modulo logarithms, in
agreement with the PQCD analysis of Belitsky, Ji, and
Yuan~\cite{bjy}.  We also shall show that the form factor ratios
obtained from the nonperturbative solutions to the Wick--Cutkosky
model~\cite{wcm,ks92} have a similar behavior.

\begin{figure}[htb]
\centering
 \includegraphics[width=0.8\textwidth]{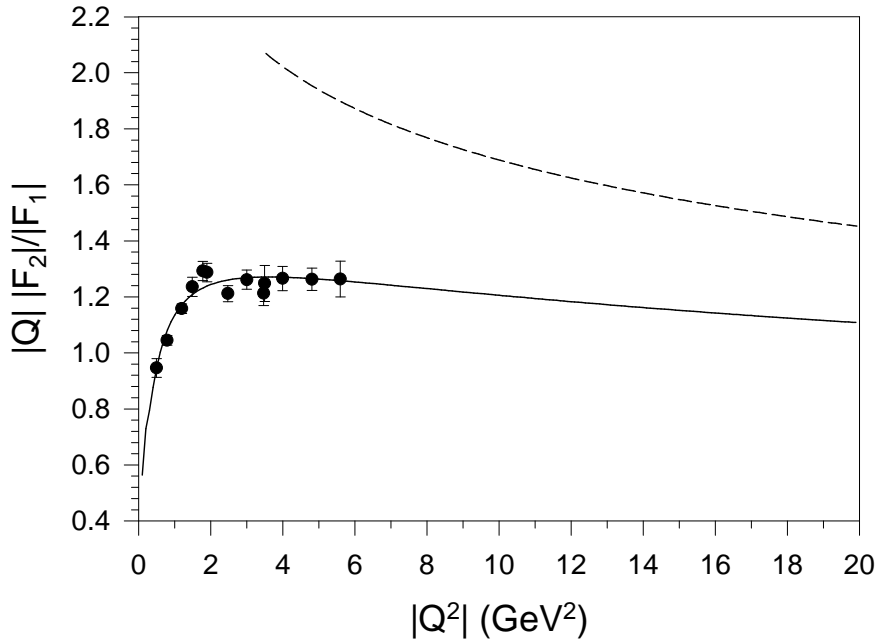}
\caption[*]{Perturbative QCD motivated fit to the Jefferson
Laboratory polarization transfer data~\cite{Jones:1999rz,Gayou:2001qd}.
The parameterization is given in Eq.~(\ref{eq:F2F1}) of the text.
The dashed line shows the
predicted form~\cite{Geshkenbein74} for timelike $q^2=-Q^2$.}
\label{fig:QF2F1}
\end{figure}

It should be noted that the analytic form predicted by
perturbative QCD is compatible with the form factor ratio
determined by the polarization transfer measurements. The detailed
analysis of baryon form factors at large $Q^2$ based on
perturbative QCD predicts the asymptotic behavior ${Q^2
F_2(Q^2)/F_1(Q^2)} \sim \log^{2+8/(9\beta)}{(Q^2/\Lambda^2)}$,
where $\beta=11-2n_f/3$~\cite{bjy}. This asymptotic logarithmic
form can be generalized to include the correct $Q^2=0$ limit and
the cut at the two-pion threshold in the timelike region.  Such a
parameterization is
\begin{equation} \label{eq:F2F1}
F_2/F_1 = \kappa_p\frac{1+(Q^2/C_1)^2 \log^{b+2}(1+Q^2/4m_\pi^2)}
               {1+(Q^2/C_2)^3 \log^b(1+Q^2/4m_\pi^2)},
\end{equation}
where for simplicity we have ignored the small factor $8/9\beta$,
as do Belitsky {\em et al.}  For the large-$Q^2$ region of the
available data, this already reduces to the asymptotic form
\begin{equation}\label{eq:F2F1a}
F_2/F_1 = \kappa_p\frac{C_2^3}{C_1^2} \frac{\log^2(Q^2/4m_\pi^2)}{Q^2}.
\end{equation}
Therefore, the values of $C_1$, $C_2$ and b are not tightly
constrained, except for the combination $C_2^3/C_1^2$. A fit to
the JLab data yields $C_1=0.79$ GeV$^2$, $C_2=0.38$ GeV$^2$, and
$b=5.1$. Thus, as shown in Fig.~\ref{fig:QF2F1}, one can fit the
form factor ratio over the entire measured range with an analytic
form compatible with the predicted perturbative QCD asymptotic
behavior.

The nominal $1 / Q^2$ power-law fall-off  of $F_2(Q^2)/F_1(Q^2)$ in
perturbative QCD is a consequence of the underlying chiral structure
of the vector interactions in QCD.  The factorized structure of hard
QCD amplitudes  predicts hadron helicity conservation~\cite{Brodsky:1981kj}
at leading twist and thus the relative suppression of the Pauli form
factor since it is a helicity-flip amplitude.  These nominal power-law
forms are also properties of dimensional counting
rules~\cite{Brodsky:1973kr,Brodsky:1974vy,Matveev:ra} for hard scattering
amplitudes in QCD.  The power-law suppression of $F_2(Q^2)/F_1(Q^2)$ is
not generally true for Yukawa theories with scalar gluons or in
quark-diquark models of the nucleon based on scalar
diquarks since the effective interactions violate chirality conservation.

Iachello, Jackson, and Lande~\cite{ijl,gk,lomon} have introduced a
model for the nucleon form factors based on dimensional counting
and perturbative QCD at high momentum plus the analytic structure
due to vector meson intermediate states.  The model gives an
excellent phenomenological description of the individual form
factors and the form factor  ratio measured  using polarization
transfer.

Although the spacelike form factors of a stable hadron are real, the
timelike form factors have a phase structure reflecting the final-state
interactions of the outgoing hadrons. The analytic structure and phases
of the form factors in the timelike regime are connected by dispersion
theory to the spacelike regime. Each of the above models predicts
a specific fall-off and phase structure of the form factors from
$ s \leftrightarrow t$ crossing to the timelike domain. As noted by
Dubnickova, Dubnicka, and Rekalo, and by Rock~\cite{d}, the phase of
the form factor ratio $G_E/G_M$ of spin-1/2 baryons in the timelike
region can be determined from measurements of the polarization of one
of the produced baryons in the exclusive process
$e^- e^+ \rightarrow B \bar B$, since the single-spin asymmetry normal
to the scattering plane requires a nonzero phase difference
between the $G_E$ and $G_M$ form factors.  As demonstrated in
Ref.~\cite{Brodsky:2003gs}, measurements of the proton polarization
in $e^+  e^- \to p \bar p$ will strongly discriminate between the
analytic forms of the models which have been suggested to fit the
proton $G_E/G_M$ data in the spacelike region.  The  single-spin
proton asymmetry is predicted to be large, of order of several
tens of percent.

The content of the remainder of the paper is as follows. The
general construction of the wave functions appearing through the
Fock decomposition of the state vector defined on the light-front
$\omega\cd x=0$ is explained in Sec.~\ref{general}. The angular
momentum properties of LFWFs are discussed in
Sec.~\ref{yukmod}. For simplicity, we  present the LFWFs for
bound states of scalar fields and for a spin-1/2 system in Yukawa
theory which can serve as a diquark-inspired model for a
three-quark hadron.  The consequences of the CC ansatz are
explored in detail. Section~\ref{LFspinors} recasts this
discussion in terms of the standard light-front Fock-state
expansion and the associated LF spinors.  This allows contact to
be made with perturbation theory. The construction of form factors
and their asymptotic behavior are described in Sec.~\ref{ff2}.
The LFWF for the valence Fock state is obtained by
integrating the Bethe-Salpeter solutions over the relative
light-front energies. The expectation that LFWFs are functions of
$M_0^2$, rather than $M_0$, is demonstrated in
Sec.~\ref{WCmodel}. The angular momentum properties of LFWFs
are illustrated using the known nonperturbative eigensolutions of
the Wick--Cutkosky model for nonzero angular momentum.
Section~\ref{3body} presents the implications for the  CC ansatz
for a three-quark nucleon state, where we show that the anomalous
asymptotic behavior of $QF_2/F_1\sim {\rm const}$ is  a
consequence of the extra constraints on the LFWFs imposed by the
CC construction. Section~\ref{conclusions} contains our
conclusions, and an appendix collects some useful definitions and
intermediate results.

\section{Light-front wave functions}
\subsection{General construction}\label{general}

The concept of a wave function of a hadron as a composite of
relativistic quarks and gluons is naturally formulated in terms of
the light-front Fock expansion at fixed light-front time,
$\sigma=x \cd \omega$~\cite{cdkm}. The null four-vector $\omega$
determines the orientation of the light-front plane; the freedom
to choose $\omega$ provides an explicitly covariant formulation of
light-front quantization. For a stationary state we can consider a
fixed light-front time and put  $\sigma=0$.

In this formulation, the eigensolution of a proton,
projected on the eigenstates of the free
Hamiltonian $ H^{QCD}_{LC}(g = 0)$, has the expansion
\begin{eqnarray}
\left\vert p, \lambda \right> &=&
\sum_{n \ge 3} \int
\psi^{\lambda}_{\lambda_1,\ldots,\lambda_n}(k_1,\ldots,k_n,p,\omega\tau)
\nonumber\\
&&\times
\delta^{(4)}\left(\sum^n_j k_j-p-\omega\tau\right) 2(\omega\cd p) d\tau
\;\prod^{n}_{i=1} \frac{d^3k_{i}}{(2\pi)^3 2\varepsilon_{k_i}}
a^{\dagger}_{\lambda_i}(\vec{k}_i)\left| 0 \right>.
\label{stv}
\end{eqnarray}
Here $a^\dagger$ is the usual creation operator and
$\varepsilon_{k_i}=\sqrt{m_i^2+\vec{k}_i^{\,2}}$. All the four-momenta are on the
corresponding mass shells: $k_j^2=m_j^2$, $p^2=M^2$, $(\omega\tau)^2=0$. The set
of light-front Fock state wave functions $
\psi^{\lambda}_{\lambda_1,\ldots,\lambda_n}(k_1,\ldots,k_n,p,\omega\tau) $
represents the ensemble of quark and gluon states possible when the proton is
intercepted at the light-front. The $\lambda_j$ label the light-front spin
projections of the quarks and gluons along the light-front quantization
direction. The scalar variable $\tau$ controls the off-shell continuation of the
wave function. From the point of view of kinematics, the four-momentum
$\omega\tau$ can be considered on equal ground with the particle four-momenta
$k_1,\ldots,k_n,p$. Being expressed through them (by squaring the equality
$\sum^n_j k_j=p+\omega\tau$), $\tau$ reads
$$
\tau=\frac{M_0^2-M^2}{2\omega\cd p},
$$
where $M_0^2=(\sum^n_j k_j)^2$. The difference $M_0-M\sim \tau$
between the effective mass $M_0$ of constituents and their bound
state mass $M$ is just a measure of the off-energy-shell effect.

If the wave functions describe a system with spin composed of
constituents with spin, they are represented through   scalar
functions multiplied by invariants constructed from the spin
matrices,  polarization vectors, and $\omega^\mu.$ In general, the
scalar functions depend on a set of scalar products of the
four-momenta $k_1,\ldots,k_n,p,\omega\tau$ with each other. One
should choose a set of independent scalar products. A convenient
way to choose these variables is the following. We define
$$
x_j=\frac{\omega\cd k_j}{\omega\cd p}, \quad
R_j=k_j-x_jp,
$$
where $\sum_{j=1}^n x_j=1$ and  $\sum_{j=1}^n R_j= \omega\tau$,
and represent the spatial part of the four-vector $R_j$ as
$\vec{R}_j=\vec{R}_{\| j} +\vec{k}_{\perp j}$, where $\vec{R}_{\|
j}$ is parallel to $\vec{\omega}$  and $\vec{k}_{\perp j}$ (with
$\sum_j^n \vec{k}_{\perp j}=0$) is orthogonal to $\vec{\omega}$.
Since $R_j\cd\omega=R_{0j}\omega_0-\vec{R}_{\|
j}\cd\vec{\omega}=0$ by definition of  $R_j$, it follows that
$R_{0 j}=|\vec{R}_{\| j}|$, and, hence, $\vec{k}^2_{\perp j}
=-R_j^2$ and $\vec{k}_{\perp i}\cd \vec{k}_{\perp j} =-R_i\cd R_j$
are expressed though the  squares and the scalar products of the
four-vectors $R_j$. Hence, they are the Lorentz and rotation
invariants. Therefore, the scalar functions should depend on
$x_j$, $\vec{k}^2_{\perp j}$ and $\vec{k}_{\perp i}\cd
\vec{k}_{\perp j}$. In terms of these variables, the integral in
Eq.~(\ref{stv}) is transformed as
\begin{eqnarray*}
\lefteqn{\int \ldots \delta^{(4)}\left(\sum^n_j k_j-p-\omega\tau\right)
    2(\omega\cd p) d\tau
\;\prod^{n}_{i=1} \frac{d^3k_{i}}{(2\pi)^{3}2\varepsilon_{k_i}}}\\
&&=
\int \ldots 2\delta\left(\sum_j^n x_j-1\right)
\delta^{(2)}\left(\sum_j^n \vec{k}_{\perp j}\right)
\prod^{n}_{i=1} \frac{d^2\vec{k}_{\perp i} dx_i}{16 \pi^3 x_i}.
\end{eqnarray*}
In this way we find that $\psi$ reduces to
$\psi^\lambda_{\lambda_1,\ldots,\lambda_n}(x_j,{\vec k_{\perp j}})$.
These wave functions  are independent of
the hadron's momentum $p^+= p\cd \omega$ and $p_\perp$, reflecting
the kinematical boost invariance of the front form. All
observables must be invariant under variation of $\omega^\mu;$  as
we shall show, this generalized rotational invariance provides an
elegant representation of angular momentum on the light front.

A scalar bound state of two spinless particles with
$p^2=(k_a+k_b)^2= M^2$ can be described using the covariant
Bethe--Salpeter function $\Phi(x_1,x_2,p)$, which in momentum
space corresponds to $\Phi_{BS}(k_a,k_b)=\Phi(q,p)$, where
$q=(k_a-k_b)/2$. For $J=0$ the Bethe--Salpeter function is a
scalar function of the Feynman virtualities with $k^\mu_a +
k^\mu_b = p^\mu$.

The explicitly covariant version of light-front dynamics~\cite{cdkm}
turns into the standard one at the particular value of
$\omega=(1,0,0,-1)$.  The conjugate direction is
defined as $\zeta$, where $\zeta \cd \omega = 2$ and
$\zeta=(1,0,0,1)$ in the standard light-front frame.
In general, the two-body scalar LFWF can only be a function of the
corresponding off-shell light-front energy
\begin{equation}\label{eq6b}
M^2_0 -M^2=(\sum_i^2 k_i \cd \zeta - p\cd \zeta )(\omega\cd
p)= 2(\omega\cd p)\tau \end{equation}
and the momentum fractions $x_i = k_i\cd \omega/ p \cd \omega$.

One can identify the corresponding two-parton LFWF by calculating
the Fourier transform of $\Phi(x_1,x_2,p)$ with the arguments
$x_1$, $x_2$ constrained to a fixed light-front time
$\sigma=\omega \cd x$, {\em i.e.}, the transform of the function
$\Phi(x_1,x_2,p)\delta(\omega\cd x_1)\delta(\omega\cd x_2)$.  In
momentum space it corresponds to integration over the minus
component of the relative momentum $k$, or in the covariant
form~\cite{cdkm}
\begin{equation} \label{bs8}
\psi =\frac{(\omega\cd k_1 )(\omega\cd k_2 )}{\pi (\omega\cd
p)}\int_{-\infty }^{+\infty }\Phi (k+\beta\omega,p)d\beta.
\end{equation}
The resulting LFWF describes a spin-zero bound state of on-shell
partons with $k^2_a=m^2_a$, $k^2_b=m^2_b$.

The main dynamical dependence of a LFWF thus involves $M^2_0 =
{(\sum k^\mu_i)}^2$, the invariant mass squared of the partons.
Since the equation of motion for the LFWF involves the off-shell
LF energy $p\cd \zeta$, the natural analytic dependence of
$\varphi$ is in the variable $M^2_0$, not $M_0$.

It is useful to examine the properties of the LFWF in the
constituent rest frame~\cite{karm80,Brodsky:1980vj}.  We will
distinguish the four vector ${\cal P}^\mu = \sum k^\mu_i$ and the
bound-state four momentum $p^\mu$.  Since there is no conservation
law for the minus components of momenta, we have $\zeta\cd p \ne
\zeta\cd {\cal P}$, and the constituent rest frame ($\vec{\cal
P}=0$) and the rest frame ($\vec{p}=0$) differ from each other. In
the constituent rest frame where $\vec {\cal P} = \sum_i \vec k_i
= \vec 0$, we can identify
 $$M_0 = {\cal P}^0= \sum_i k^0_i =
\sum_i \sqrt{{\vec k}^2_i+m^2_i}.$$  It is also convenient to make
an identification of the front-form and the usual instant-form wave
functions in this frame.

In general, LFWFs are eigenstates of the LF angular momentum
operator~\cite{cdkm}
\begin{equation}\label{ac1}
\vec{J} = -i[\vec{k}\times \partial/\partial\vec{k}\,]-i[\vec{n}\times
\partial/\partial\vec{n}] +\frac{1}{2}\vec{\sigma},
\end{equation}
where $\vec n$ is the spatial component of $\omega$ in the
constituent rest frame ($\vec {\cal P}=\vec 0$).  Although this form
is written specifically in the constituent rest frame, it can be
generalized to an arbitrary frame by a Lorentz boost.

Normally the generators of angular rotations in the LF formalism
contain interactions, as in the Pauli--Lubanski formulation;
however, the LF angular momentum operator can also be represented
in the kinematical form (\ref{ac1}) without interactions.  The key
term is the generator of rotations of the LF plane
$-i[\vec{n}\times\partial/\partial\vec{n}]$ which replaces the
interaction term; it appears only in the explicitly covariant
formulation, where the dependence on $\vec{n}$ is present. Details
can be found in~\cite{cdkm}. The application of the  LF angular
momentum operator (\ref{ac1}) to the scalar function
$\varphi(M^2_0,x)$ verifies that it describes a $J=0$ state.

\subsection{The Yukawa model}\label{yukmod}

The general form of the LFWF of a spin-1/2 system
composed of spin-half and spin-zero constituents is given by
\begin{equation}\label{eq6}
\psi(k_1,p)=\bar{u}(k_1)\left(\varphi_1+ \frac{M\hat{\omega}} {\omega\cd
p}\varphi_2\right)u(p),
\end{equation}
where $\omega=(\omega_0,\vec{\omega})$ is the four-vector determining the
orientation of the light-front plane $\omega\cd x=0$.
Here
$\hat{\omega}=\omega_{\mu}\gamma^{\mu}$, and $\bar{u}(k_1)$ is the
conventional Dirac spinor of the spin-half constituent, and
$u(p)$ is the spinor of the bound state.

The general wave function (\ref{eq6}) in Yukawa theory  is
determined by two scalar components $\varphi_1$ and $\varphi_2$,
each a function of $M^2_0$ and $x$.  In contrast, the wave
function corresponding to the CC ansatz used in~\cite{mf} is
equivalent\footnote{Equivalence has been established~\cite{karm98}
in the sense that equivalent results are obtained for form
factors.} to the form
\begin{equation}\label{eq6a}
\psi=c_k c_p\bar{u}(k_1)\Lambda_+ u(p) f_1,
\end{equation}
where ${\cal P}=k_1+k_2$ is the sum of the constituent momenta,
$M_0=\sqrt{{\cal P}^2}$, $c_k=1/\sqrt{m+\varepsilon_k}$,
$c_p=1/\sqrt{M+\varepsilon_p}$,
$\varepsilon_k=\sqrt{m^2+\vec{k}^2}$, and
$\varepsilon_p=\sqrt{M^2+\vec{p}\,^2}$
(for simplicity we assume that constituents have equal masses $m_a=m_b=m$).
The matrix $\Lambda_+$ in
(\ref{eq6a}) is the projection operator
\begin{equation}\label{eq4}
\Lambda_+=\frac{\hat{\cal P}+M_0}{2 M_0},
\end{equation}
which in the constituent rest frame becomes
\begin{equation}\label{eq4z}
\Lambda_+=\left(
\begin{array}{ll}
1 & 0\\0 & 0
\end{array}
\right).
\end{equation}
Hence, in the constituent rest frame we obtain
\begin{equation}\label{spinors}
c_k\bar{u}(k_1)\Lambda_+ =\chi_1^\dagger(1,\; 0), \quad c_p\Lambda_+ u(p)=\left(
\begin{array}{l} 1\\ 0
\end{array}\right)\chi_N,
\end{equation}
with $\chi_1$ and $\chi_N$ being two-component spinors for constituent 1 and the
nucleon, respectively.  From (\ref{eq6a}) we reproduce the analog of the
three-quark CC ansatz used in Ref.~\cite{mf}, which is
\begin{equation}\label{eq1a}
\psi_{\sigma_1}^{\sigma}= \chi^{\dagger\sigma_1} \chi^\sigma f_1
=\delta^{\sigma_1\sigma} f_1.
\end{equation}
This should be contrasted with the general form (in the constituent rest frame)
\begin{equation}\label{eq1}
\psi_{\sigma_1}^{\sigma}=\chi^{\dagger\sigma_1}\left(f_1
+\frac{i}{k}[\vec{n}\times \vec{k}]\cd \vec{\sigma} \;f_2\right)\chi^{\sigma},
\end{equation}
where $\vec{k}=\vec{k}_1=-\vec{k}_2$ and $\vec{n}=\vec{\omega}/|\vec{\omega}|$ in
this frame. The additional $f_2$ term represents a separate dynamical
contribution with orbital angular momentum, to be contrasted with the purely
kinematical contributions of orbital angular momentum from Melosh rotations. The
form (\ref{eq1}) can be found by substituting into Eq.~(\ref{eq6}) the explicit
form of the Bjorken--Drell (BD) spinors, as given in the constituent frame by
Eq.~(\ref{spa1}) of the appendix. The two functions $f_1$ and $f_2$ are then
determined in terms of $\varphi_1$ and $\varphi_2$ as
\begin{eqnarray}\label{eq11}
f_1&=&\frac{c_k c_p}{2M_0}(M_0+m)\left[m(M_0+m)+M_0(xM+(1-x)M_0)\right] \varphi_1
\nonumber\\
&& + \frac{c_k c_p}{M_0}M(M_0+m)\left(xM_0+m\right) \varphi_2,
\nonumber\\
f_2&=&\frac{c_kc_p}{M_0} kM(M_0+M) \varphi_2 - \frac{c_k c_p}{2M_0}k(M_0^2-M^2)
\varphi_1,
\end{eqnarray}
with $x=x_1=1/2-\vec{k}\cd\vec{n}/M_0$. The inverse relations read
\begin{eqnarray}\label{eq12}
\varphi_1&=&\frac{c_k c_p}{2M_0}(M_0+M) f_1-
 \frac{c_k c_p}{2M_0 k}(M_0+M)\left(xM_0+m\right)
f_2,
\nonumber\\
\varphi_2&=&\frac{c_k c_p}{4M_0 M}(M_0^2-M^2)  f_1 \nonumber\\
&& + \frac{c_k c_p}{4M_0 M k}(M_0+M)\left[m(M_0+m)+M_0(xM+(1-x)M_0)\right] f_2 \
.
\end{eqnarray}
Note that the wave functions $\varphi_1$ and $\varphi_2$ are Lorentz scalars and
analytic functions of $M_0^2$ and $x$.  The derived amplitudes $f_1$ and $f_2$
contain kinematic factors linear in $M_0$ which arise from the reduction of the
covariant form to the Pauli spinor form in the constituent rest frame. The
application of the LF angular momentum operator (\ref{ac1}) to the LFWFs of the
Yukawa theory in Eq.~(\ref{eq1}) verifies that these wave functions are in fact
states with $J=1/2$, $J_z=\pm 1/2$.

In the non-relativistic limit, where $M=2m$, $M_0=2m$, and $x=1/2$, we get
\begin{equation}
\varphi_1=\frac{1}{2\sqrt{2}m}f_1-\frac{1}{k\sqrt{2}}f_2,\quad
\varphi_2=\frac{1}{k\sqrt{2}}f_2. \label{nonrelreln}
\end{equation}
The component $f_2$ is of relativistic origin.  In the
non-relativistic limit $c\to \infty$ (here $c$ is the speed of
light), the LF plane $t+z/c=0$ turns into $t=0$.  Any $\vec{n}$
dependence in (\ref{eq1}) should then disappear, and this happens
if $f_2$ becomes negligible.

We see that in order to obtain the constrained CC form
(\ref{eq1a}) from the general form (\ref{eq6}), one should
eliminate, according to (\ref{spinors}), the second components of
spinors and neglect the second component $f_2$ in (\ref{eq1}).
This is natural in a non-relativistic approximation, when both the
second components of spinors and the dependence on the light-front
orientation indeed disappear.  One can use this form of the wave
function to estimate the influence of relativistic effects on the
static nucleon properties~\cite{cc}.  However, there is no
compelling reason to use the form (\ref{eq1a}) in the asymptotic
relativistic domain.

The form (\ref{eq1a}), determined by the single component $f_1$, implies a
relation between $\varphi_1$ and $\varphi_2$ in (\ref{eq6}).  Setting
$f_2$ to zero in (\ref{eq12}), we obtain
\begin{eqnarray}\label{eq12c}
\varphi_1&=&\frac{c_k c_p}{2M_0}(M_0+M) f_1,
\nonumber\\
\varphi_2&=&\frac{c_k c_p}{4M_0 M}(M_0^2-M^2)  f_1,
\end{eqnarray}
and, hence,
\begin{equation}\label{eq17d}
\varphi_2=\frac{M_0-M}{2M}\varphi_1.
\end{equation}
Since $M_0$ is large for large momenta, we see that in the asymptotic
regime the component $\varphi_2$ dominates: $\varphi_2\approx
M_0\varphi_1/(2M)$.

Thus the ansatz (\ref{eq1a}) is equivalent to an assumption that
the component $\varphi_2$ dominates in (\ref{eq6}).  We shall show
below that this predominance of $\varphi_2$ over $\varphi_1$,
generated by the wave function (\ref{eq1a}), results in the
$QF_2/F_1\to {\rm const}$ asymptotic behavior.

\section{Light-front spinors}\label{LFspinors}

The discussion of the previous section can also be given in terms
of a LF Fock-state expansion and the associated LF spinors. The
LFWF of a hadron with spin projection $J_z = \pm {1\over 2}$ is
represented by the function
$\psi^{J_z}_{\lambda_1,\ldots,\lambda_n}(x_i,{\vec k}_{\perp i})$,
where
\begin{equation}
k_i=(k^+_i,k^-_i,{\vec k}_{i \perp})= \left(x_i P^+, \frac{{\vec
k}_{\perp i}^2+m_i^2}{x_i P^+}, {\vec k}_{i\perp}\right)
\end{equation}
specifies the momentum of each constituent and $\lambda_i$ specifies its
light-front helicity in the $z$ direction.  The light-front fractions $x_i
= k^+_i/P^+$ are positive and satisfy $\sum_i x_i = 1$.  We note that
$M_0^2 = {\sum^n_{i=1} {k^2_{\perp i}+m^2_i\over x_i}}=(\sum_i k_i)^2$ is
Lorentz invariant, and the scalar part of the LFWF is a function
of only $x_i$ and $M_0^2$.

For a spin-1/2 state with two constituents in Yukawa theory, we write
$\psi_\lambda^{J_z}(x,\vec{k}_\perp)
\equiv\psi_{\lambda}^{J_z}(x_i,\vec{k}_{\perp i})$, where $\lambda=\lambda_1$ is
the helicity of the fermion,  $x=x_1$, and $\vec{k}_\perp=\vec{k}_{1\perp}$.
(We use the subscript 1 for the fermion and 2 for the scalar constituent.) The
two-particle Fock state with total momentum $(P^+,\vec{P}_\perp)$ and spin $J_z$
is then given by
\begin{equation}\label{sn1}
\left|P^+, \vec P_\perp = \vec 0_\perp,J_z\right>=
\int\frac{dx d^2 k_{\perp}}{16\pi^3\sqrt{x(1-x)}}\sum_\lambda
\psi_\lambda^{J_z}(x,\vec{k}_\perp)|xP^+,\vec{k}_\perp,\lambda\rangle.
\end{equation}
The Fock-state ket on the right is defined by
\begin{equation}
|xP^+,\vec{k}_\perp,\lambda\rangle\equiv
|k_1^+=xP^+,k_2^+=(1-x)P^+;
  \vec{k}_{1\perp}=\vec{k}_\perp,\vec{k}_{2\perp}=-\vec{k}_\perp;
  \lambda_1=\lambda\rangle
\end{equation}
and normalized by
\begin{equation}
\langle k'_i{}^+, {\vec k\,'_{i\perp}}, \lambda' | k_i^+, {\vec
k_{i\perp}},\lambda\rangle = \prod_{i=1}^2 16\pi^3 k_i^+ \delta(k_i^{\prime +}
-k_i^+) \delta( {\vec k\,'_{i\perp}} - {\vec k_{i\perp}})
\delta_{\lambda',\lambda}. \label{normalize}
\end{equation}

The four functions $\psi_{\pm 1/2}^{\uparrow}$ and
$\psi_{\pm 1/2}^{\downarrow}$ provide a representation of
the LFWFs for $J_z=\uparrow,\downarrow$~\cite{bhms}.  The associated
spinors are the LF spinors $u^{LF}$~\cite{BL80,Brodsky:1997de},
which in the hadron rest frame ($\vec{p}=0$) are given by
\begin{eqnarray}\label{eq16a}
\bar{u}^{LF\sigma_1}(k_1)&=&\frac{1}{\sqrt{2(k_{10}+k_{1z})}}
  \chi^{\dagger\sigma_1}
\left(m+k_{10}+\sigma_z(\vec{\sigma}\cd \vec{k}_1)
-(k_{10}-m)\sigma_z-\vec{\sigma}\cd \vec{k}_1\right), \nonumber\\
u^{LF\sigma}(p)&=&\frac{1}{\sqrt{2M}} \left(\begin{array}{c}
2M\\0\end{array}\right)\chi^{\sigma}.
\end{eqnarray}
The general form (\ref{eq6}) has a simple form in the light-front
spinor representation. A straightforward calculation
gives\footnote{The wave function (\ref{eq6}) and the one given
here are related as follows:
$\psi {\rm (in\ (\ref{eq6}))}
=\sqrt{x(1-x)}\psi {\rm (here)}$. In
this expression, $k_x$ and $k_y$ are equivalent to $k^1$ and
$k^2$, and $J_z$ and $L_z$ denote $J^3$ and $L^3$.}

\begin{eqnarray}\label{eq17}
\sqrt{1-x}\;\psi^{\uparrow}_{+\frac{1}{2}}(x,\vec{k}_\perp)&=&
\left(M+\frac{m}{x}\right)\varphi_1+2M\varphi_2,
\nonumber\\
\sqrt{1-x}\;\psi^{\uparrow}_{-\frac{1}{2}}(x,\vec{k}_\perp)&=&
-\frac{(+k_x+ik_y)}{x}\varphi_1,
\nonumber\\
\sqrt{1-x}\;\psi^{\downarrow}_{+\frac{1}{2}}(x,\vec{k}_\perp)&=&
\frac{(+k_x-ik_y)}{x}\varphi_1,
\nonumber\\
\sqrt{1-x}\;\psi^{\downarrow}_{-\frac{1}{2}}(x,\vec{k}_\perp)&=&
\left(M+\frac{m}{x}\right)\varphi_1+2M\varphi_2.
\end{eqnarray}

Note that in the perturbative Yukawa model\footnote{In the
case of perturbative models, a single-particle wave function
$\psi^{J_z}_{\lambda_1} = \sqrt{Z} \delta^{2}(\vec k_\perp) \delta(1-x)
\delta_{J_z \lambda_1}$
is present, where the normalization constant
$Z$ ensures unit probability.  The perturbative Yukawa model
wave functions can be formally differentiated with respect to the boson
mass in order to simulate the fall-off of the wave function of a composite
hadron and eliminate the single-particle Fock component.}
one obtains~\cite{bhms}
\begin{equation}
\varphi_1=\frac{g}{M^2-M^2_0},\quad \varphi_2=0.
\label{yukawa1}
\end{equation}
In this way we reproduce the
wave functions (44) and (46) of~\cite{bhms}.  The general solution
in (\ref{eq17}), which does not require any assumptions,
differs from the perturbative solution only by the
contribution $2M\varphi_2$ in the components
$\psi^{\uparrow}_{+\frac{1}{2}}$ and $\psi^{\downarrow}_{-\frac{1}{2}}$.

In terms of the LF spinor representation, the CC ansatz (\ref{eq1a})
implies particular forms for the components of (\ref{eq17}).
Substituting the expression (\ref{eq17d}) for $\varphi_2$ into
Eqs.~(\ref{eq17}), we find
\begin{eqnarray}\label{eq17c}
\sqrt{1-x}\;\psi^{\uparrow}_{+\frac{1}{2}}(x,\vec{k}_\perp)&=&
\left(M_0+\frac{m}{x}\right)\varphi_1,
\nonumber\\
\sqrt{1-x}\;\psi^{\uparrow}_{-\frac{1}{2}}(x,\vec{k}_\perp)&=&
-\frac{(+k_x+ik_y)}{x}\varphi_1,
\nonumber\\
\sqrt{1-x}\;\psi^{\downarrow}_{+\frac{1}{2}}(x,\vec{k}_\perp)&=&
\frac{(+k_x-ik_y)}{x}\varphi_1,
\nonumber\\
\sqrt{1-x}\;\psi^{\downarrow}_{-\frac{1}{2}}(x,\vec{k}_\perp)&=&
\left(M_0+\frac{m}{x}\right)\varphi_1.
\end{eqnarray}%
We see that the CC ansatz in (\ref{eq1a}) is equivalent to the replacement $M\to
M_0$ in the perturbative LF components given in~\cite{bhms}.  The anomalous
dependence on $M_0$ is the source of the discrepancy of the ansatz with the
asymptotic behavior of the components with different angular momentum projections
found in~\cite{jmy} from a perturbative model based on the iteration of the
one-gluon exchange kernel. The perturbative QCD counting rules for hard
scattering exclusive amplitudes~\cite{Brodsky:1973kr,Brodsky:1974vy,Matveev:ra}
and the fall-off of hadronic LFWFs~\cite{jmy} at high transverse momentum can
also be derived~\cite{Polchinski:2001tt,Brodsky:2003px} without the use of
perturbation theory using the conformal properties of the AdS/CFT correspondence
between gauge theory and string theory~\cite{Maldacena:1997re}.

The component $\psi^{\uparrow}_{-\frac{1}{2}}$ corresponds to
$L_z=1$, and the component $\psi^{\uparrow}_{+\frac{1}{2}}$
corresponds to $L_z=0$.  The PQCD analysis based on the exchange
of a gluon gives~\cite{jmy}
\begin{equation}\label{rat}
\psi ({L_z=1})/\psi ({L_z=0})=
\psi^{\uparrow}_{-\frac{1}{2}}/\psi^{\uparrow}_{+\frac{1}{2}} \sim
k_\perp \varphi_1/\varphi_2 \sim
1/k_\perp,
\end{equation}
which implies $Q^2F_2/F_1={\rm const}$~\cite{jmy}. We confirm this
in the next section. On the other hand, (\ref{eq17d}) or
(\ref{eq17c}), corresponding to the CC ansatz, gives
\begin{equation}\label{rat1}
\psi ({L_z=1})/\psi ({L_z=0})=
\psi^{\uparrow}_{-\frac{1}{2}}/\psi^{\uparrow}_{+\frac{1}{2}} \sim
k_\perp \varphi_1/\varphi_2 \sim {\rm const.}
\end{equation}
In turn, the ratio (\ref{rat1}) results in the asymptotic behavior
$QF_2/F_1={\rm const}$, as we will see in the next section.

\boldmath
\section{Form factors and the $M_0$ dependence of light-front
wave functions}\label{ff2}\unboldmath

\subsection{Dirac and Pauli form factors}

In the case of a spin-${1\over 2}$ composite system, the Dirac and
Pauli form factors $F_1(q^2)$ and $F_2(q^2)$ are defined by
\begin{equation}
      \langle P'| J^\mu (0) |P\rangle
       = \bar u(P')\, \Big[\, F_1(q^2)\gamma^\mu +
F_2(q^2){i\over 2M}\sigma^{\mu\alpha}q_\alpha\, \Big] \, u(P),
\label{Drell1}
\end{equation}
where $q^\mu = (P' -P)^\mu$, $u(P)$ is the bound-state spinor, and
$M$ is the mass of the composite system. In the light-front
formalism it is convenient to identify the Dirac and Pauli form
factors from the helicity-conserving and helicity-flip vector
current matrix elements of the $J^+$ current component~\cite{BD80}
\begin{equation}
\VEV{P+q,\uparrow\left|\frac{J^+(0)}{2P^+}
\right|P,\uparrow} =F_1(q^2) ,
\label{BD1}
\end{equation}
\begin{equation}
\VEV{P+q,\uparrow\left|\frac{J^+(0)}{2P^+}\right|P,\downarrow}
=-(q^1-{\mathrm i} q^2){F_2(q^2)\over 2M}.
\label{BD2}
\end{equation}
We use the standard light-front frame
($q^{\pm}=q^0\pm q^3$) where
\begin{eqnarray}
q &=& (q^+,q^-,{\vec q}_{\perp}) = \left(0, \frac{-q^2}{P^+},
{\vec q}_{\perp}\right), \nonumber \\
P &=& (P^+,P^-,{\vec P}_{\perp}) = \left(P^+, \frac{M^2}{P^+},
{\vec 0}_{\perp}\right),
\label{LCF}
\end{eqnarray}
and $q^2=-2 P \cd q= -{\vec q}_{\perp}\,^2$ is the square of
the four-momentum transferred by the photon.

Using Eqs.~(\ref{BD1}) and (\ref{sn1}) we have
\begin{equation}
F_1(q^2) =
\int\frac{{\mathrm d}^2 {\vec k}_{\perp} {\mathrm d} x }{16 \pi^3}
\Big[\psi^{\uparrow\ *}_{+\frac{1}{2}}(x,{\vec k'}_{\perp})
\psi^{\uparrow}_{+\frac{1}{2}}(x,{\vec k}_{\perp})
+\psi^{\uparrow\ *}_{-\frac{1}{2}}(x,{\vec k'}_{\perp})
\psi^{\uparrow}_{-\frac{1}{2}}(x,{\vec k}_{\perp})\Big],
\label{BDF1as}
\end{equation}
where
\begin{equation}
{\vec k'}_{\perp}={\vec k}_{\perp}+(1-x){\vec q}_{\perp}.
\label{BDF1b}
\end{equation}
{}From Eqs.~(\ref{BD2}) and (\ref{sn1}), we have
\begin{equation}
F_2(q^2)
={-2M\over (q_x-{\mathrm i}q_y)}
\int\frac{{\mathrm d}^2 {\vec k}_{\perp} {\mathrm d} x }{16 \pi^3}
\Big[\psi^{\uparrow\ *}_{+\frac{1}{2}}(x,{\vec k'}_{\perp})
\psi^{\downarrow}_{+\frac{1}{2}}(x,{\vec k}_{\perp})
+\psi^{\uparrow\ *}_{-\frac{1}{2}}(x,{\vec k'}_{\perp})
\psi^{\downarrow}_{-\frac{1}{2}}(x,{\vec k}_{\perp})
\Big].
\label{BDF1bz}
\end{equation}
The individual wave functions are given by (\ref{eq17}); substitution yields
\begin{eqnarray}
\lefteqn{F_1(q^2)=
\frac{1}{16\pi^3}\int \frac{d^2 k_\perp dx}{x^2(1-x)}}&&
\label{f1a}\\ &&\times\left\{
\left[x(2mM+xM^2)+m^2+k^2_{\perp}-\frac{1}{4}(1-x)^2 Q^2
\right]\varphi'_1\varphi_1\right. \nonumber\\
&&+\left.2Mx(m+xM)(\varphi'_1\varphi_2 + \varphi'_2\varphi_1)
+4M^2x^2\varphi'_2\varphi_2\right\} , \nonumber\\
\lefteqn{F_2(q^2)\ =\ \frac{M}{8\pi^3}\int \frac{d^2 k_\perp
dx}{x^2(1-x)}}&& \label{f2a}\\
&&\times\left\{(1-x)\left(m+xM\right)
\varphi'_1\varphi_1-2Mx\frac{\vec{k}_{\perp}\cd {\vec q}_{\perp}
}{Q^2} \displaystyle{(\varphi_1\varphi'_2
-\varphi_2\varphi'_1)}\right. \nonumber\\ &&\left.+
Mx(1-x)\displaystyle{(\varphi_1\varphi'_2
+\varphi_2\varphi'_1)}\right\}, \nonumber
\end{eqnarray}
where the $\vec{k}_{\perp}$ variable has been shifted by
$-\frac{1}{2}(1-x)\vec{q}_{\perp}$, so that
\begin{equation}
\varphi_{1,2}=\varphi_{1,2}(x, \vec{k}_\perp
     -\frac{1}{2}(1-x){\vec q}_{\perp}),\quad
\varphi'_{1,2}=\varphi_{1,2}(x,
    \vec{k}_\perp+\frac{1}{2}(1-x){\vec q}_{\perp}).
\label{p12p}
\end{equation}

The required parton invariant masses $M_0$ and ${M'}_0$ are
\begin{equation}
M_0^2=\frac{\left(\vec{k}_{\perp} -\frac{1}{2}(1-x){\vec
q}_{\perp}\right)^2+m^2}{x(1-x)}, \quad
{M'}_0^2=\frac{\left(\vec{k}_{\perp}+\frac{1}{2}(1-x){\vec
q}_{\perp}\right)^2+m^2}{x(1-x)} . \label{m0m0p}
\end{equation}
Being expressed in terms of the invariant masses, the form factor
integrands contain squares of masses: $M_0^2$ and ${M'}_0^2$.

One can see that if $\varphi_2$ is smaller than $\varphi_1$ or of
the order of $\varphi_1$ for large ${\vec k}_{\perp}^2$, the
leading terms in $F_1$ are $ k^2_{\perp}$ and $-\frac{1}{4}(1-x)^2
Q^2$. An analytical calculation of the asymptotic behavior of the
form factors with the power-law wave function
$\varphi_1=N/(M_0^2+\beta^2)^{n}$ shows that these two leading
contributions cancel each other.  Thus the leading term becomes
$\sim \log Q^2$ instead of $\sim Q^2$ and the ratio $F_2/F_1$
becomes $\sim 1/\log(Q^2/m^2)$ .  The same result is found for
pseudoscalar coupling, {\em i.e.}, with the wave function which is
obtained by inserting in (\ref{eq6}) the matrix $\gamma_5$.  As
explained in the introduction, the cancellation of the leading
term in the Dirac form factor in the scalar gluon models is
related to the violation of chirality conservation.

We summarize in Table \ref{tab1} how the asymptotic behavior of the form
factor ratio depends on the asymptotic properties of the LFWFs in the
Yukawa model.  Since the scalar part of the LFWF is a function
of $M_0^2$ and $x_i$, and $\psi (L_z=\pm 1)$ contains the $k_x\pm ik_y$
prefactor, $\psi (L_z=1) / \psi (L_z=0)$ can only be an odd
power of $k_{\perp}$.  The third and fourth columns of Table 1 correspond
to the cases of (\ref{rat1}) and (\ref{rat}), respectively.
\begin{table}[h!]
\begin{center}
\begin{tabular}{cccccc}
\hline\hline
\\[-1.5ex]
 $\frac{\psi(L_z=1)}{\psi(L_z=0)}$ & $k^2_{\perp}$ &  $k_{\perp}$ &
const &$\frac{1}{k_{\perp}}$ & $\frac{1}{k^2_{\perp}}$\\[1ex]
\hline
\\[-1.5ex]
$\frac{F_2}{F_1}$ & $\frac{1}{Q^3}$ &   $\frac{1}{Q^2}$ &  $\frac{1}{Q}$ &
$\frac{1}{Q^2}$ &  $\frac{1}{Q^3}$\\[1ex]
\hline\hline
\end{tabular}
\end{center}
\caption{The dependence of the asymptotic form of the form factor ratio on
the asymptotic behavior of the LF components
$\psi(L_z=1)/\psi(L_z=0)=\psi^{\uparrow}_{-\frac{1}{2}}/
\psi^{\uparrow}_{+\frac{1}{2}}$ in the Yukawa model. \label{tab1}}
\end{table}

The cancellation of the leading power-law contribution is specific to the
scalar and pseudoscalar diquark models.  In the case of a spin-1/2 system
comprising a spin-1/2 quark and a spin-1 diquark, the hadron wave function
is determined in general by six independent components.  To see the effect
coming from the spin-1 diquark, consider a wave function in
the one-component form
\begin{equation}\label{eq6v}
\psi_{\sigma_1 , \lambda}(k_1,k_2,p)=e_{\nu}^{*(\lambda)}(k_2)
\bar{u}_{\sigma_1}(k_1)\gamma^{\nu}u(p)\varphi_1.
\end{equation}
Here $e_{\nu}^{(\lambda)}(k_2)$ is the spin-1 polarization vector. The sum
over polarizations results in the propagator
$(g_{\nu\nu'}-k_{2\nu}k_{2\nu'}/\mu^2)$ in the form
factor calculation. In this model we find for the form factor
ratio $$ \frac{F_2}{F_1}\sim
\frac{\log(Q^2/m^2)}{Q^2}, $$ which is close to the fit (\ref{eq:F2F1a}),
differing only by a factor $\log(Q^2/m^2)$. The same perturbative
behavior (up to a coefficient) also occurs in the asymptotic behavior of
the electron form factor in QED. If a scalar or pseudoscalar
coupling is also present, the vector contribution will dominate the
asymptotic behavior. We emphasize that for all three couplings, both
form factors decrease as  an integral power of
$Q^2$ (modulo powers of $ \log(Q^2/m^2)$), not as a power of $Q$.
 As shown in Fig.~\ref{fig:QF2F1}, a fit to the
form factor ratio based on powers of $Q^2$ and powers of $\log
Q^2$ describes the experimental data well.

\subsection{Consequences of the CC constraint}

The CC ansatz $f_2=0$ is equivalent to a quark--scalar-diquark model,
but with the additional condition (\ref{eq17d}).  As we have seen
this introduces anomalous terms in the
LFWFs which are linear in $M_0$. Substituting $\varphi_2$
from (\ref{eq17d}) into Eqs.~(\ref{f1a}), (\ref{f2a}), we obtain:
\begin{eqnarray}\label{f1b}
F_1(q^2)&=&\frac{1}{32\pi^3}\int \frac{d^2k_\perp dx}{x^2(1-x)}
\left[x(1-x)(M_0^2+{M'}^2_0)+2x^2M_0M'_0+2xm(M_0+M'_0)\right.\nonumber\\
&-&\left.(1-x)^2Q^2\right] \varphi'_1\varphi_1
\\
 F_2(q^2)& =& \frac{M}{16\pi^3}\int
\frac{d^2k_\perp dx}{x^2(1-x)}
\left[2(1-x)m+x(1-x)(M_0+M'_0)\right. \nonumber\\
&-&\left.\frac{x^2}{Q^2}(M_0-M'_0)^2(M_0+M'_0)\right]
\varphi'_1\varphi_1 \label{f2b}
\end{eqnarray}
The same result is of course obtained by direct calculation  with
the wave function (\ref{eq6a}).

Because of the terms with $M_0^2$, ${M'}^2_0$, and $M_0M'_0$, the
form factor $F_1$ contains the second power  $Q^2$ relative to the
term $2(1-x)m$ in $F_2$, which does not contain $M_0$ or $M'_0$.
This occurs independently of whether the cancellation between
$k^2_{\perp}$ and $-\frac{1}{4}(1-x)^2 Q^2$ takes place or not.
If it does not take place, we get an extra contribution to the
$Q^2$ power.  This only changes the coefficient of $Q^2$.
Similarly, because of the term with $(M_0+M'_0)$, $F_2$ contains
the first power of $Q$.  This extra factor of $M_0$ (and, hence,
$Q$) results in the nominal asymptotic behavior
\begin{equation}\label{mf1}
\frac{QF_2}{F_1}={\rm const.}
\end{equation}
We emphasize that this behavior follows from the CC ansatz
(\ref{eq1a}) (or $f_2=0$ in (\ref{eq1})), which in turn, is
equivalent to the relation (\ref{eq17d}).  After the constraint is
imposed, the asymptotic behavior (\ref{mf1}) follows without
further dynamical assumptions.

\begin{figure}[htb]
\centering
 \includegraphics[width=0.8\textwidth]{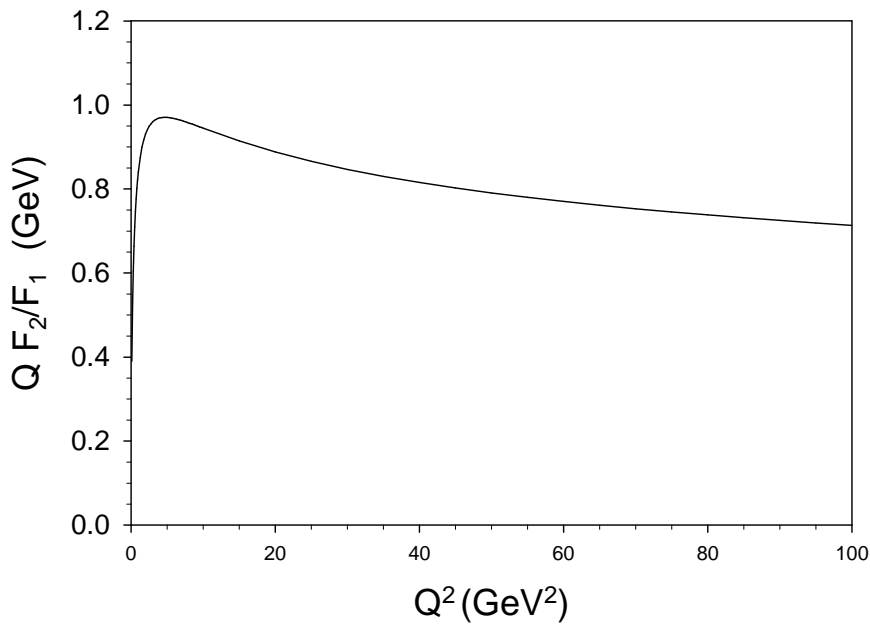}
\caption[*]{Asymptotic behavior of the form factor ratio
$QF_2/F_1$ as determined from Eqs.~(\ref{f1a}) and (\ref{f2a}) of the text
with $\varphi_1\propto(M_0^2+(0.6\,\mbox{GeV})^2)^{-3.5}$ and $\varphi_2$
fixed by Eq.~(\ref{eq17d}).  The constituent mass is $m=0.3$ GeV,
and the bound-state mass is $M=0.94$ GeV.  The choice of $\varphi_2$
corresponds to the Chung--Coester ansatz and nominally implies a
constant asymptotic behavior for the ratio, but here is well fit by
$Q^{-0.25}$.}
\label{fig:asymptotic}
\end{figure}

The  coefficient functions $\varphi_i$ are in general functions of
$M^2_0$ since in the equations of motion only the quantity $M_0^2$
appears.   In the next section this is shown explicitly in the
case of the Wick--Cutkosky model. Explicit calculation of the form
factors, with $\varphi_1=N/(M_0^2+\beta^2)^{3.5}$, as assumed in
Ref.~\cite{mf}, and with $\varphi_2$ fixed by Eq.~(\ref{eq17d}),
qualitatively replicates the nonconstant, nonintegral asymptotic
behavior found in~\cite{mf}. Figure~\ref{fig:asymptotic} shows the
result for $QF_2/F_1$, which is approximately constant for a small
range of intermediate $Q^2$ but is well fit by $Q^{-0.25}$ for
large $Q^2$. Similar results are obtained with an exponential wave
function.

The arguments by Ralston {\em et al.}~\cite{Ralston} and
Kroll~\cite{Kroll}, in favor of $QF_2/F_1$ being asymptotically
constant, reduce to a discussion of the following ratio of wave
function matrix elements:
\begin{equation}
\frac{F_2}{F_1}=\frac{\langle \bar{\psi}_1\psi_0\rangle}
                   {Q\langle b\bar{\psi}_0\psi_0\rangle}.
\end{equation}
(See for example Eq.~(5) of Ref.~\cite{Ralston} and Eq.~(11) of
Ref.~\cite{Kroll}.) They arrive at this ratio by including an
$L_z=1$ wave function in the contribution to $F_2$, so that there
is an overlap with the $L_z=0$ wave function.  However, the
$L_z=1$ wave function $\psi_{-1/2}^\uparrow$ also contributes to
$F_1$ in an overlap with itself, as given above in
Eq.~(\ref{BDF1as}).  This introduces an additional term of order
$Q^2$ in the denominator and makes $Q^2F_2/F_1$ asymptotically
constant.

\boldmath
\section{The Wick--Cutkosky model and $M_0^2$ dependence}\label{WCmodel}
\unboldmath

The Wick--Cutkosky model is based on the ladder approximation to
the Bethe--Salpeter (BS) equation in $\phi^2\chi$ field theory.
The LFWFs in this model for the $J=L=0$ and $J=L=1$ bound states
have been computed in~\cite{karm80}.  We shall show below that if
the wave function is represented in four-dimensional form, the
scalar components depend on $M_0^2$ as expected on general
grounds.  Indeed, for $J=1$ the BS function reads~\cite{wcm,nak69}
\begin{eqnarray}\label{bs1}
\Phi_{1\lambda}(q,p)&=&
\tilde{q}Y^*_{1\lambda}\left(\frac{\tilde{\vec{q}}}
{\tilde{q}}\right)\Phi_1(q,p),\\
\Phi_1(q,p)&=& -i\int_{-1}^{+1}\frac{
g_1(z,M)dz}{(m^2-M^2/4-q^2-zp\cd q-i\epsilon)^4}, \nonumber
\end{eqnarray}
where $g_1(z,M)$ is the spectral function satisfying a specific
differential equation~\cite{wcm,nak69}. We will not need the
explicit form of $g_1$. Here $q=(k_a-k_b)/2$,
$\tilde{\vec{q}}=L^{-1}(\vec{v})\vec{q}$ has the sense of relative
momentum in the rest frame given by $\vec{p}=0$, and
$L^{-1}(\vec{v})$ is the Lorentz boost with
$\vec{v}=\vec{p}/\varepsilon_p$. In an arbitrary frame the
spherical function in (\ref{bs1}) can be replaced by
\begin{equation}\label{y}
\tilde{q}Y^*_{1\lambda}\left(\frac{\tilde{\vec{q}}}{\tilde{q}}\right)\to
-\sqrt{\frac{3}{4\pi}} e_{\mu}^{(\lambda)}(p)q^{\mu},
\end{equation}
where $e_{\mu}^{(\lambda)}(p)$ is the spin-1 polarization vector.

Using the relation (\ref{bs8}) between the LFWF and the BS function and Eq.
(\ref{y}), we find the LFWF in the form
\begin{equation}\label{lfinv}
\psi_{1\lambda}=-\sqrt{\frac{3}{4\pi}}
e_{\mu}^{(\lambda)}(p)\psi^{\mu},\quad
\psi^{\mu}=k^{\mu}\varphi_1+\frac{\omega^{\mu}}{\omega\cd
p}\varphi_2,
\end{equation}
where
\begin{eqnarray}\label{phis}
\varphi_1&=& \phantom{-}\frac{2g_1(1-2x,M)}{3(M^2_0-M^2)^3x^2(1-x)^2},
\nonumber\\
\varphi_2&=& -\frac{g_1(1-2x,M)}
{6(M^2_0-M^2)^2x^2(1-x)^2}
-\frac{g'_1(1-2x,M)}{3(M^2_0-M^2)^2x(1-x)}.
\end{eqnarray}
The prime in $g'_1(z,M)$ means differentiation with respect to $z$.
We see that the components $\varphi_1$ and $\varphi_2$ depend on $M_0^2$, as
expected.  A similar result holds for the $L=0$ state:
\begin{equation}\label{bs10}
\psi =\frac{g(1-2x,M)}{2\sqrt{\pi }x(1-x) (M^2_0-M^2)^2},
\end{equation}
where $g(z,M)$ is the corresponding spectral function. It does not
depend linearly on $M_0$.

The representation (\ref{lfinv}) for the Wick--Cutkosky wave
function is analogous to the representation (\ref{eq6}) for the
Yukawa model. These results for the LFWFs of the bound states of
the Wick--Cutkosky model can also be obtained by computing the
instant-form wave function and boosting to infinite momentum, as
in Weinberg's $P_z \to \infty$ method~\cite{Weinberg66}.

Note that the LFWF (\ref{lfinv}) in the constituent rest frame
has the form
\begin{equation}\label{wf1}
\psi_{1\lambda}(\vec{k},\vec{n})=f_1(\vec{k}\,^2,\vec{n}\cd \vec{k})
kY^*_{1\lambda}\left(\frac{\vec{k}}{k}\right) + f_2(\vec{k}\,^2,\vec{n}\cd
\vec{k}) Y^*_{1\lambda}\left(\vec{n}\right),
\end{equation}
where the scalar components $f_{1,2}$ can be expressed in terms of
$\varphi_{1,2}$ from (\ref{lfinv}). This representation is
analogous to the representation (\ref{eq1}) for the Yukawa model.
Since the BS function (\ref{bs1}) describes a state with  angular
momentum $J=1$, the corresponding LFWF (\ref{wf1}), which is
derived from the BS function without any approximation, definitely
has the same angular momentum.  It is an eigenfunction of the
angular momentum operator (\ref{ac1}) (omitting the spin operator
$\frac{1}{2}\vec{\sigma}$).

\section{Consequences of the CC ansatz for three-body light-front
wave functions}\label{3body}

The simplified three-quark  LFWF of the nucleons introduced by
Coester and Chung, and used in~\cite{mf} for a model of the proton
form factors, has the same spin structure as the
quark--scalar-diquark model---two quarks form a spin-zero diquark,
and the spin of nucleon is determined by the spin of the third quark.
The wave function is symmetrized relative to the permutations of the quarks
according to $SU(6)$ flavor-spin symmetry.  For example, in the
constituent rest frame, the Pauli-spinor representation of the
three-quark wave function corresponding to the state of zero spin
and zero isospin of the quark pair $1 2$ is
$$
\psi(12,3)=
\chi^{(S=0)}(1,2)\;\chi^\dagger_{\sigma_3}\delta^{\sigma_3\sigma} \cd
\xi^{(I=0)}(1,2)\; \xi^\dagger_{\tau_3}\delta^{\tau_3\tau},
$$
where
$\chi^{(S=0)}(1,2)=\chi^\dagger_{\sigma_1}i\sigma_y\chi^\dagger_{\sigma_2}$
is the spin-zero wave function of the two quarks and
$\xi^{(I=0)}(1,2)=
\xi^\dagger_{\tau_1}i\sigma_y\xi^\dagger_{\tau_2}$ is the
isospin-zero wave function. The nucleon spin-isospin wave function
is obtained by symmetrization:
$$
\psi(1,2,3)=\psi(12,3)+\psi(23,1)+\psi(31,2).
$$
Using the Fierz identities, we may transform it to the form
\begin{equation}\label{eq27}
\psi=\frac{1}{\sqrt{72}} \psi_S[3+(\vec{\sigma}_{12}\cd\vec{\sigma}_{3N})
(\vec{\tau}_{12}\cd\vec{\tau}_{3N})].
\end{equation}
This form is totally symmetric with respect to spin and isospin.
We have introduced here the symmetric momentum-dependent part
$\psi_S$.  The factors in (\ref{eq27}) should be understood as
$$
1\equiv (\chi^\dagger_{\sigma_1}i\sigma_y\chi^\dagger_{\sigma_2})
\;(\chi^\dagger_{\sigma_3}\chi_{\sigma_N}), \quad (\vec{\sigma}_{12}\cd
\vec{\sigma}_{3N})\equiv (\chi^\dagger_{\sigma_1}\vec{\sigma}\sigma_y
\chi^\dagger_{\sigma_2})
\;(\chi^\dagger_{\sigma_3}\vec{\sigma}\chi_{\sigma_N}),
$$
and similarly for the isospin part.
The wave function (\ref{eq27}) is just the
ansatz used in~\cite{mf}.

In the covariant Yukawa model, the general form of the two-parton
LFWF (\ref{eq1}) contains two components, in contrast to the
one-component CC ansatz (\ref{eq1a}).  The general form of the
three-quark nucleon wave function contains sixteen
components~\cite{karm98} in contrast to the one-component ansatz
(\ref{eq27}).  Thus the CC ansatz (\ref{eq27}) is even a stronger
constraint  than in the two-body Yukawa model.

As in the two-body case (\ref{eq6a}), the three-body wave function
(\ref{eq27}) can be recast in the four-dimensional
form~\cite{karm98}\begin{eqnarray}\label{eqcc}
\psi&=&\frac{\psi_S}{\sqrt{72}}c_1 c_2 c_3 c_N\left\{3[\bar{u}(k_1)\Lambda_+
\gamma_5 U_c\bar{u}(k_2)]\;[\bar{u}(k_3)\Lambda_+ u_N(p)]\right.
 \nonumber\\
 &&
 \left.-[\bar{u}(k_1)\Lambda_+\gamma^\mu\Lambda_-
U_c\bar{u}(k_2)]\;[\bar{u}(k_3)\Lambda_+\gamma_\mu\gamma_5 \Lambda_+
u_N(p)](\vec{\tau}_{12}\cd\vec{\tau}_{3N})\right\},\end{eqnarray} where
$\Lambda_-=\frac{\hat{\cal P}-M_0}{2 M_0}$ and $U_c=\gamma_2\gamma_0$
is the charge conjugation matrix. The coefficients are
$c_1=1/\sqrt{m+\varepsilon_{k_1}}$, etc. The
projection operator $\Lambda_+$ is again given by Eq.~(\ref{eq4});
however, the two-body constituent momentum is replaced by the
three-body momentum
$$
{\cal P}=k_1+k_2+k_3,\quad M_0^2={\cal P}^2.
$$
In the constituent rest frame, where $\vec{\cal P}=0$, we have
${\cal P}_0=M_0$, and the wave function (\ref{eqcc}) reduces to
(\ref{eq27}).

The second term in (\ref{eqcc}) does not contribute to the proton
form factor~\cite{karm98,mf}.  The first term is factorized.  We
suppose that the photon interacts with the third quark.  The first
factor $[\bar{u}(k_1)\Lambda_+ \gamma_5 U_c\bar{u}(k_2)]$ is the
same for $F_1$ and $F_2$, and, therefore, it does not change their
ratio.  Since only the second factor $[\bar{u}(k_3)\Lambda_+
u_N(p)]$ gives different contributions to $F_1$ and $F_2$, we
rewrite the wave function in a factorized form
\begin{equation}\label{eqsh}
\psi\propto \psi_S\;\bar{u}(k_3)\Lambda_+ u_N(p).
\end{equation}
This coincides with the CC-constrained Yukawa wave function (\ref{eq6a}).

The form factor calculation using the three-quark wave function
determined by (\ref{eq27}) and (\ref{eqsh}) differs from the
two-body calculation in Sec.~\ref{ff2} only in the use of
three-body kinematics.  The result is
\begin{eqnarray}\label{f1c}
F_1(q^2)&=&\frac{1}{2}\int
\left[x_3(1-x_3)(M_0^2+{M'}^2_0)+2x_3^2M_0M'_0+2x_3m(M_0+M'_0)
-(1-x_3)^2Q^2\right]
 \nonumber\\
&&\qquad\qquad\qquad\times G(1,2,3,Q^2) D,
\\
 F_2(q^2)& =&M\int
\left[2(1-x_3)m+x_3(1-x_3)(M_0+M'_0)
-\frac{x_3^2}{Q^2}(M_0-M'_0)^2(M_0+M'_0)\right] \nonumber\\
&&\qquad\qquad\qquad\times G(1,2,3,Q^2) D. \label{f2c}
\end{eqnarray}
Here  $D$ is the three-body phase volume, and $G(1,2,3,Q^2)$ is a
function of the variables of all three quarks and of $Q^2$.  We do
not need the explicit form of $G$.  $M_0$ is the effective
three-body mass, which depends on
$\vec{R}_{3\perp}-\frac{1}{2}(1-x_3)\vec{q}_{\perp}$, whereas
$M'_0$ depends on
$\vec{R}_{3\perp}+\frac{1}{2}(1-x_3)\vec{q}_{\perp}$. Because of
factorization of the wave function (\ref{eqsh}), these formulas
(found by direct calculation) coincide, after evident changes of
notation, with the corresponding expressions (\ref{f1b}) and
(\ref{f2b}) for the two-body form factors for the CC ansatz.

At $Q^2\equiv \vec{q}_{\perp}^{\,2}\to \infty$, both $M'_0$ and $M_0$
tend to $Q$.  For the leading term we find
\begin{eqnarray}\label{cc4}
F_1&\propto & \ldots Q^2,
\\
 F_2&\propto &\ldots Q,
 \nonumber
\end{eqnarray}
which reproduce the ratio $QF_1/F_2={\rm const}$. As before, this
asymptotic behavior is a consequence of the specific constraints
on the LFWFs.

On the other hand, we have mentioned in Sec.~\ref{ff2} that
scalar and pseudoscalar couplings  result in the asymptotic ratio
$F_2/F_1 \sim 1/\log(Q^2/m^2)$ and vector coupling in $F_2/F_1
\sim \log(Q^2/m^2)/Q^2$.  These couplings are not really related
to a hypothesis of predominance of diquarks, but simply represent
different spin structures of the nucleon wave function. The total
number of these structures is sixteen~\cite{karm98}.  Some of
these other structures, in addition to the vector one, may also
contribute to the asymptotic behavior $F_2/F_1 \sim
\log(Q^2/m^2)/Q^2$.

\section{Conclusions}\label{conclusions}

The explicitly Lorentz-invariant formulation of the front form provides
a general method for determining the general structure of light-front
wave functions.  We have also used the fact that the angular momentum
of a bound state can be defined covariantly within the Bethe-Salpeter
formalism:  The LFWFs for an $n$-particle Fock state evaluated at equal
light-front time $\sigma = \omega\cd x$ can be obtained by integrating
the covariant Bethe-Salpeter functions over the corresponding relative
light-front energies.  The resulting LFWFs  are eigenstates of the
kinematic angular momentum operator (\ref{ac1}). The result is that
the LFWFs are functions of the invariant mass squared of the constituents
$M^2_0= (\sum k^\mu)^2$ and the light-cone momentum fractions $x_i= {k_i\cd
\omega / p \cd \omega},$ each multiplying spin-vector and polarization
invariants involving $\omega^\mu$, where $\omega=(\omega_0,\vec{\omega})$
is the four-vector determining the orientation of the light-front plane
$\omega\cd x=0$.

We have presented the structure of LFWFs for two and
three-particle bound states using the explicitly Lorentz-invariant
formulation of the front form~\cite{cdkm}.  As examples we have
given the explicit form of the LFWFs for spin-$0$ and spin-$1$
eigenstates of the nonperturbative eigensolutions of the
Wick--Cutkosky model, as well as examples of spin-$1/2$ states
constructed using perturbation theory. For example, the LFWF of a
spin-1/2 system composed of spin-half and spin-zero constituents
has the general form
$$\psi(k_1,p)=\bar{u}(k_1)\left(\varphi_1+ \frac{M\hat{\omega}} {\omega\cd
p}\varphi_2\right)u(p),$$
where the $\varphi_i$ are functions of the square of the
invariant mass and the light-front momentum fractions. The
orbital angular momentum prefactor in the constituent rest
frame is proportional to $\vec{\omega}\times \vec {k}\cd\vec{S}$.

An important test of the LF computations is light-front invariance---although the
LFWFs depend on the choice of the light-front quantization direction, all
observables, such as matrix elements of local current operators, form factors,
and cross sections, must be independent of $\omega^\mu.$  We have computed the
large momentum transfer behavior of the ratio of Pauli and Dirac form factors of
the nucleon using the exact relation for spacelike current matrix elements in
terms of LFWFs. The dependence of the invariant mass squared implies that hadron
form factors computed from the overlap integrals of LFWFs are analytic functions
of $Q^2$ in agreement with dispersion theory, the PQCD analysis of Belitsky, Ji,
and Yuan~\cite{bjy}, and conformal arguments~\cite{Brodsky:2003px}, as well as
with the form factor ratios obtained using the nonperturbative solutions to the
Wick--Cutkosky model. We have also shown that a fit to the Pauli to Dirac form
factor ratio incorporating the predicted perturbative QCD $1/Q^2$ and $\log{Q^2}$
asymptotic dependence describes the recent Jefferson laboratory polarization
transfer data well. In contrast, we have shown that the LFWFs introduced by Chung
and Coester to parameterize the static and low-momentum properties of the
nucleons correspond to the spin-locking of a quark with the spin of its parent
nucleon, together with a positive-energy projection constraint.  These extra
constraints lead to an anomalous linear dependence of the LFWFs on the invariant
mass of the constituents and an anomalous dependence of form factors on $Q$
rather than $Q^2.$

The CC construction of relativistic light-front wave functions was
introduced to represent the properties of the nucleons in the low
momentum transfer domain. However, there are a number of difficulties
with extending the CC form to the high momentum transfer domain:

(1) If one applies the CC ansatz to a bound state of a spin-half
quark and a scalar (the Yukawa model), the quark is constrained to
have the same spin projection as the bound state nucleon, $S^z_q =
S^z_p$, when one uses the conventional BD spinor representation.
Thus the only orbital angular momentum allowed by the CC
constraint is the kinematical angular momentum arising from the
lower components of the BD spinor.  The spin-locked CC constraint
does not allow for the full degrees of freedom of a relativistic
system.

(2) If one compares the CC ansatz to wave functions generated in
perturbation theory, the net result is to replace the bound-state
mass $M$ in the numerator of the LFWFs by the invariant mass of
the constituents $M_0 = \sqrt{\sum^n_{i=1} {k^2_{\perp
i}+m^2_i\over x_i}}$.  This replacement leads to the anomalous
growth of the CC LFWF at large transverse momentum.  For example,
if one applies the CC ansatz to the QCD quark splitting function,
the ultraviolet $\log Q^2$ behavior of $q(x,Q^2)$ will derive from
two sources: the standard behavior in $k_\perp$ arising from
perturbative QCD plus the presence of a term $M^2_0 \propto
k^2_\perp$ due to the CC ansatz.  The presence of the latter
source would destroy DGLAP evolution in $Q^2$.

(3) The LFWFs of the $J=0$ and $J=1$ bound states can be obtained
explicitly in the Wick--Cutkosky model.  The form factor ratios of
the spin-1 system obtained in this non-perturbative analysis is
given by a quadratic $Q^2$ dependence and by $\log
Q^2$~\cite{ks92}.  The application of the CC ansatz leads to terms
in the wave function which are linear in $M_0$ rather than the
quadratic dependence of the explicit solutions.

(4) In the case of the simple Yukawa theory, the effective
interaction in the CC model has the form $H_I^{\rm eff} = g
{\bar{\psi}}_q \Lambda \psi \phi$ where $\Lambda$ is a non-local
positive-energy projection operator.  The presence of the
projection operator $\Lambda$ conflicts with the usual relations
obtained from crossing and particle--antiparticle symmetry. For
example, consider the electroproduction amplitude $\gamma^* p \to
q + qq$,  where, for simplicity, the $qq$ diquark can be taken as
the Yukawa scalar $\phi$.  The Born amplitude with quark exchange
in the $t$ channel has the form
$$M^\mu_{\gamma^* p \to q \phi}(p \cd q,t,q^2) \propto
{e_q \overline u \gamma^\mu u \over t-m^2_q }\times
\psi(x,k_\perp)$$ where
$$t-m^2_q = x(M^2-M^2_0),$$
and $x= x_{Bj} = {-q^2\over 2 p \cd q}$.  The CC ansatz introduces a linear
term in $M_0$ in the electroproduction amplitude.  If we now use $s \to t$
crossing to obtain the process $\gamma^* \bar q \to \bar p \phi$, the
presence of a linear term in $M_0$ in the LFWF gives a contribution to the
amplitude $M_{\gamma^* \bar q \to \bar p \phi}(s,t,q^2)$ which is
proportional to $\sqrt s$ at fixed momentum transfer $t$.  This anomalous
Regge behavior corresponds to fermion exchange in the $t$ channel, which,
however, is not present in this amplitude.

Again, we emphasize that these difficulties concern the
extrapolation of the CC ansatz to the asymptotic region.  These
problems do not appear in the original work~\cite{cc} where the CC
ansatz was only applied to static nucleon properties and to form
factors at relatively small momentum transfer.

Light-front wave functions are the fundamental amplitudes which
relate hadrons to their fundamental quark and gluon degrees of
freedom. We have shown how one can exhibit the general analytic
structure of light-front wave functions, including states with
nonzero orbital angular momentum. A key element of this analysis
is the use of the explicitly Lorentz-invariant formulation of the
front form where the normal to the light-front is specified by a
general null vector $\omega^\mu.$ The resulting LFWFs are
functions $\psi_n(x_i,k_{\perp i})$ of the invariant mass squared
of the constituents $M^2_0= (\sum k^\mu)^2$ and the light-cone
momentum fractions $x_i= {k_i\cd \omega / p \cd \omega}$, which
multiply invariant prefactors constructed from the spin-matrices,
polarization vectors, and $\omega^\mu$ in the case of nonzero
orbital angular momentum. The LFWFs corresponding to definite
total angular momentum are eigenstates of a kinematic angular
momentum operator and satisfy all Lorentz symmetries of the front
form, including boost invariance. We have illustrated these
properties using known nonperturbative eigensolutions of the
Wick--Cutkosky model for nonzero angular momentum.  The dependence
of LFWFs on the invariant mass squared implies that current matrix
elements and hadron form factors are analytic functions of $Q^2$
in agreement with dispersion theory and perturbative QCD. We have
also shown that a model incorporating this analytic property and
leading-twist perturbative QCD constraints is consistent with
recent data for the ratio of proton Pauli and Dirac form factors
determined by the polarization transfer method.

\section*{Acknowledgments}

This work was supported in part by the Department of Energy, contracts
DE-AC03-76SF00515 and DE-AC05-84ER40150 (S.J.B.), DE-FG02-98ER41087 (J.R.H.), by
the LG Yonam Foundation (D.S.H.), and by the Exchange Visitor Program P-1-00162
(V.A.K.).  One of the authors (V.A.K.) is sincerely grateful for the warm
hospitality of the Theory Group at SLAC where part of this work was performed. We
also thank Jerry Miller and Carl Carlson for helpful discussions.

\section*{Appendix}\label{appen}

We use the convention $a^{\pm}=a^0{\pm}a^3$, which gives $a\cd b={1\over
2}(a^+b^- + a^-b^+) -{\vec a}_{\perp}\cd {\vec b}_{\perp}$, and use the
$\gamma$ matrices in the Dirac representation
\begin{equation}
\gamma^0=\left(
\begin{array}{cc}
1&0\\
0&-1
\end{array}
\right) ,\ \ \vec{\gamma}=\left(
\begin{array}{cc}
0&\vec{{\sigma}}\\
-\vec{\sigma}&0
\end{array}
\right) . \label{gammamtx}
\end{equation}
A caret indicates the inner product of a four-vector with the $\gamma$
matrices, so that $\hat{k}=k_\mu \gamma^\mu$.

The solutions of $\bar{u}(k_1){\hat{k}_1}=m\bar{u}(k_1)$ and
${\hat{p}}u(p)=Mu(p)$, in terms of the BD spinors, are given by
\begin{eqnarray}\label{spa}
\bar{u}^{\sigma_1}(k_1)&=&u^{\dagger\sigma_1}(k_1)\gamma^0=
\sqrt{\varepsilon_{k_1}+m}\chi^{\dagger\sigma_1}
\left(1,\ -\frac{\vec{\sigma}\cd\vec{k_1}}{\varepsilon_{k_1}+m}\right),
\nonumber\\
 u^{\sigma}(p)&=& \sqrt{\varepsilon_p+M} \left(\begin{array}{c}
1\\
\frac{\displaystyle\vec{\sigma}\cd\vec{p}} {\displaystyle\varepsilon_p+M}
\end{array}\right) \chi^{\sigma},
\end{eqnarray}
where $\chi^{\sigma}$ is a two-component spinor, $\varepsilon_k =
\sqrt{\vec{k}\,^2+m^2}$, and $\varepsilon_p = \sqrt{\vec{p}\,^2+M^2}$.
In the constituent rest frame,
where $\vec{p}+\vec{\omega}\tau=\vec{k}_1+\vec{k}_2=0$,
we introduce the variables $\vec{k}$, $\vec{n}$, $\vec{k}_1\equiv \vec{k}$,
and $\vec{\omega}=\vec{n}\omega_0$.  We also find
$$
\vec{p}=-\vec{\omega}\tau=-\vec{n}\frac{M_0^2-M^2}{2M_0},\quad
\varepsilon_p=\frac{M_0^2+M^2}{2M_0}.
$$
The value of $\tau$ here was obtained by squaring the equality
$p+\omega\tau=k_1+k_2$, which gives $\tau=(M_0^2-M^2)/(2\omega\cd p)$, and
by using $\omega\cd p=\omega\cd (k_1+k_2)=\omega_0 M_0$.  In this way, we
find the BD spinors in the constituent rest frame to be written as
\begin{eqnarray}\label{spa1}
\bar{u}^{\sigma_1}(k_1)&=& \sqrt{\varepsilon_{k}+m}\chi^{\dagger\sigma_1}
\left(1,-\frac{\vec{\sigma}\cd\vec{k}}{\varepsilon_{k}+m}\right),
\nonumber\\
 u^{\sigma}(p)&=& \frac{1}{\sqrt{2M_0}} \left(\begin{array}{c}
\phantom{-}M_0+M\\
-(M_0-M)\;\vec{\sigma}\cd\vec{n}
\end{array}\right) \chi^{\sigma}.
\end{eqnarray}

The factor $\frac{M\hat{\omega}}{\omega\cd p}$ in (\ref{eq6}) is
transformed as
$$
\frac{M\hat{\omega}}{\omega\cd
p}=\frac{M}{M_0}(\gamma_0-\vec{n}\cd\vec{\gamma}) = \frac{M}{M_0}\left(
\begin{array}{cc}
1&-\vec{n}\cd\vec{\sigma}\\
\vec{n}\cd\vec{\sigma}&-1
\end{array}
\right).
$$
We also introduce
$$
x=\frac{\omega\cd k_1}{\omega\cd
p}=\frac{1}{2}\left(1-\frac{\vec{n}\cd\vec{k}}{\varepsilon_k}\right)=
\frac{1}{2}-\frac{\vec{n}\cd\vec{k}}{M_0}.$$
Substituting these expressions into Eq.~(\ref{eq6}), we reproduce the
wave function (\ref{eq1}) with the components $f_1,f_2$ given by
Eqs.~(\ref{eq11}).

The light-front spinor is defined as~\cite{BL80}
\begin{equation}\label{eq16c}
u^{LF\sigma}(k)=
{1\over {\sqrt{2(k^0+k^3)}}}\left(\begin{array}{l}
k^0+m+\vec{\sigma}\cd\vec{k}\sigma^3
\\
(k^0-m)\sigma^3+\vec{\sigma}\cd\vec{k}
\end{array}
\right)\chi^{\sigma},
\end{equation}
which gives
\begin{equation}\label{eq16}
\bar{u}^{LF\sigma_1}(k_1)=\frac{1}{\sqrt{2(k_1^0+k_1^3)}}
\chi^{\dagger\sigma_1}
\left(k_1^0+m+\sigma^3(\vec{\sigma}\cd \vec{k}_1) ,
\ -(k_1^0-m)\sigma^3-\vec{\sigma}\cd \vec{k}_1\right).
\end{equation}
The BD and LF spinors are connected by the following unitary relations (at
$\vec{n}||z$):
\begin{equation}
\left(
\begin{array}{c}
u^{+{1\over 2}}(p)\\
u^{-{1\over 2}}(p)
\end{array}
\right)= {1\over {\sqrt{2p^+(p^0+M)}}} \left(
\begin{array}{cc}
(p^++M)&-p^R\\
p^L&(p^++M)
\end{array}
\right) \left(
\begin{array}{c}
u^{LF\, +{1\over 2}}(p)\\
u^{LF\, -{1\over 2}}(p)
\end{array}
\right), \label{s6}
\end{equation}
\begin{equation}
\left(
\begin{array}{c}
u^{LF\, +{1\over 2}}(p)\\
u^{LF\, -{1\over 2}}(p)
\end{array}
\right)= {1\over {\sqrt{2p^+(p^0+M)}}} \left(
\begin{array}{cc}
(p^++M)&p^R\\
-p^L&(p^++M)
\end{array}
\right) \left(
\begin{array}{c}
u^{+{1\over 2}}(p)\\
u^{-{1\over 2}}(p)
\end{array}
\right). \label{s7}
\end{equation}


\end{document}